\documentclass[sigconf]{acmart}
\setcopyright{none}
\usepackage{caption}
\usepackage{stfloats}
\usepackage{enumitem}
\usepackage{graphicx} 
\usepackage{subcaption}
\usepackage{siunitx}
\usepackage{tikz}
\usepackage[pages=some]{background}
\AtBeginDocument{%
  }

\copyrightyear{2026}
\acmYear{2026}
\setcopyright{cc}
\setcctype{by}
\acmConference[MLCAD '26]{2026 ACM/IEEE International Symposium on Machine Learning for CAD}{September 07--09, 2026}{Jeju Island, Republic of Korea}
\acmBooktitle{2026 ACM/IEEE International Symposium on Machine Learning for CAD (MLCAD '26), September 07--09, 2026, Jeju Island, Republic of Korea}
\acmDOI{10.1145/3831599.3840318}
\acmISBN{979-8-4007-2878-5/2026/09}

\newcommand{\sysname}{PRISM}

\author{Haocheng Xu}
\email{haochx5@uci.edu}
\affiliation{\institution{University of California, Irvine}
\city{Irvine}\state{California}\country{USA}}

\author{Ye Qiao}
\email{yeq6@uci.edu}
\affiliation{\institution{University of California, Irvine}
\city{Irvine}\state{California}\country{USA}}

\author{Phyo Pyae Moe Aung}
\email{ppaung@uci.edu}
\affiliation{\institution{University of California, Irvine}
\city{Irvine}\state{California}\country{USA}}

\author{Alok Mishra}
\email{alok.mishra@hpe.com}
\affiliation{\institution{HPE Labs}
\city{Milpitas}\state{California}\country{USA}}

\author{Pavana Prakash}
\email{prakash@hpe.com}
\affiliation{\institution{HPE Labs}
\city{Milpitas}\state{California}\country{USA}}

\author{Rolando Pablo Hong Enriquez}
\email{rhong@hpe.com}
\affiliation{\institution{HPE Labs}
\city{Milpitas}\state{California}\country{USA}}

\author{Adam Han Wu}
\email{adamhw@uci.edu}
\affiliation{\institution{University of California, Irvine}
\city{Irvine}\state{California}\country{USA}}

\author{Zhiheng Chen}
\email{zhihenc5@uci.edu}
\affiliation{\institution{University of California, Irvine}
\city{Irvine}\state{California}\country{USA}}

\author{Dejan Milojicic}
\email{dejan.milojicic@hpe.com}
\affiliation{\institution{HPE Labs}
\city{Milpitas}\state{California}\country{USA}}

\author{Sitao Huang}
\email{sitaoh@uci.edu}
\affiliation{\institution{University of California, Irvine}
\city{Irvine}\state{California}\country{USA}}

\renewcommand{\shortauthors}{Haocheng Xu et al.}
\begin{document}

\title[Structure-augmented LLMs for HLS Pragma Optimization]{Structure-augmented LLMs \\for High-Level Synthesis Pragma Optimization}

\begin{abstract}
Pragma insertion drives the quality of high-level synthesis (HLS) designs. Choosing the right directives demands expert knowledge and reasoning about loop nesting, data dependences, and memory layout.
While existing large language models (LLMs) show promise in code generation, they lack explicit program-structure awareness, limiting their ability to suggest effective pragmas.
We present 
\sysname{}, 
a novel structure-augmented LLM that closes this gap by adding compiler-grade structural reasoning to a pretrained, frozen code LLM.
It combines three hierarchical program representations, Abstract Syntax Tree (AST), Control-Flow Graph (CFG), and Data-Flow Graph (DFG), injecting them into a specific transformer layer while keeping original code tokens in a separate stream. The cross-attention gate at the injection point allows falling back to the pretrained representation when its structural signal is unhelpful.
On zero-shot evaluation in HLS-Eval, \sysname{} synthesizes 3.5× as many kernels as Llama3-8B (26.9\% vs. 7.7\%), and on the kernels where it does succeed, it produces designs that are 2.31× faster (geomean) than GPT-5-mini's.
In the agentic flow, the \sysname{} codegen outperforms other baselines when optimizing complex code and drives the average normalized improvement across the HLS-Eval suite to \SI{26.4}{\percent}.

\end{abstract}

\maketitle
\backgroundsetup{opacity=1, scale=1, angle=0, contents={
\begin{tikzpicture}[remember picture, overlay]
\node[anchor=north east, inner xsep=50pt, inner ysep=10pt] at (current page.north east) {
\href{https://www.acm.org/publications/policies/artifact-review-and-badging-current}{
\includegraphics[width=50pt]{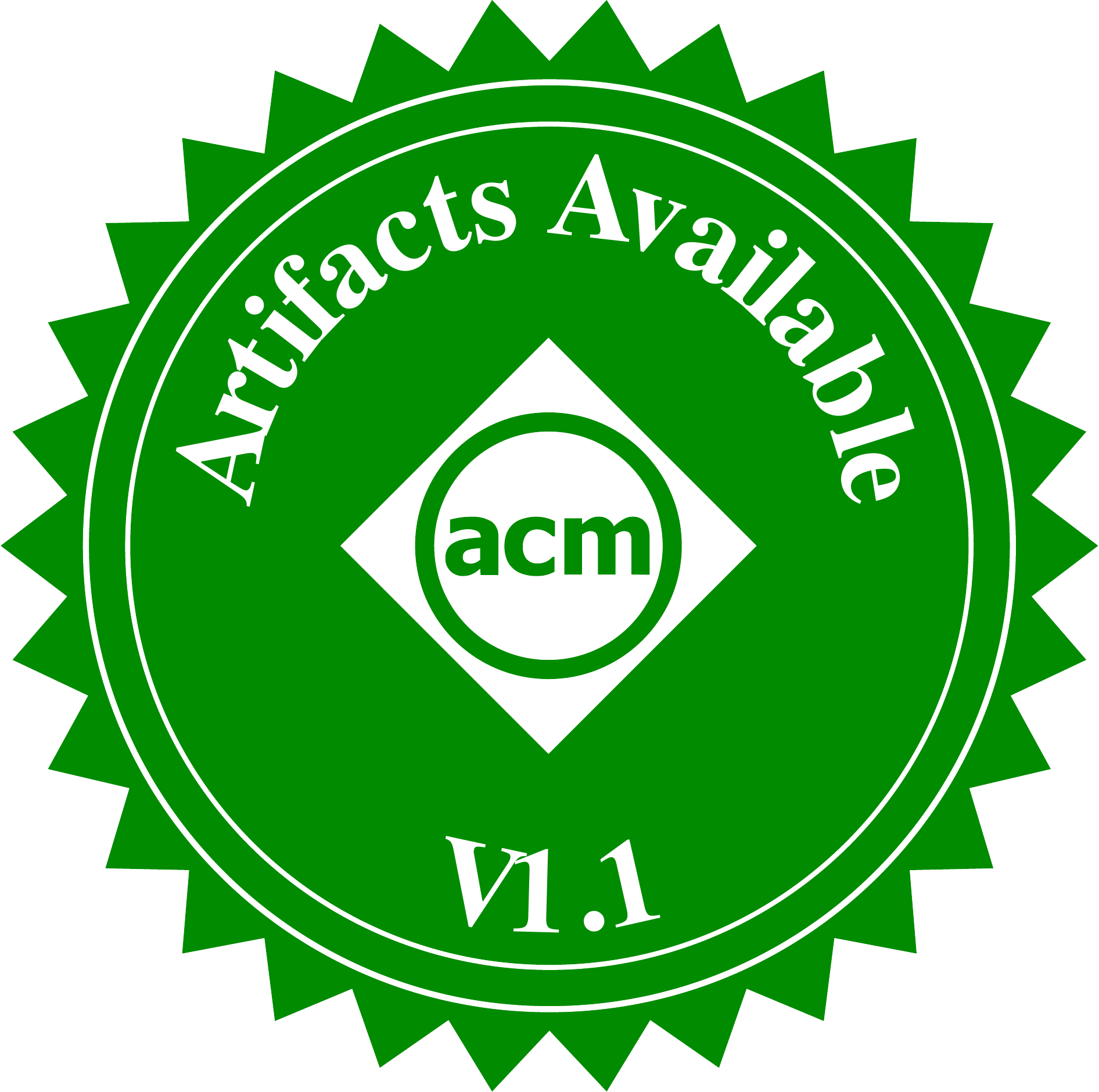}
\includegraphics[width=50pt]{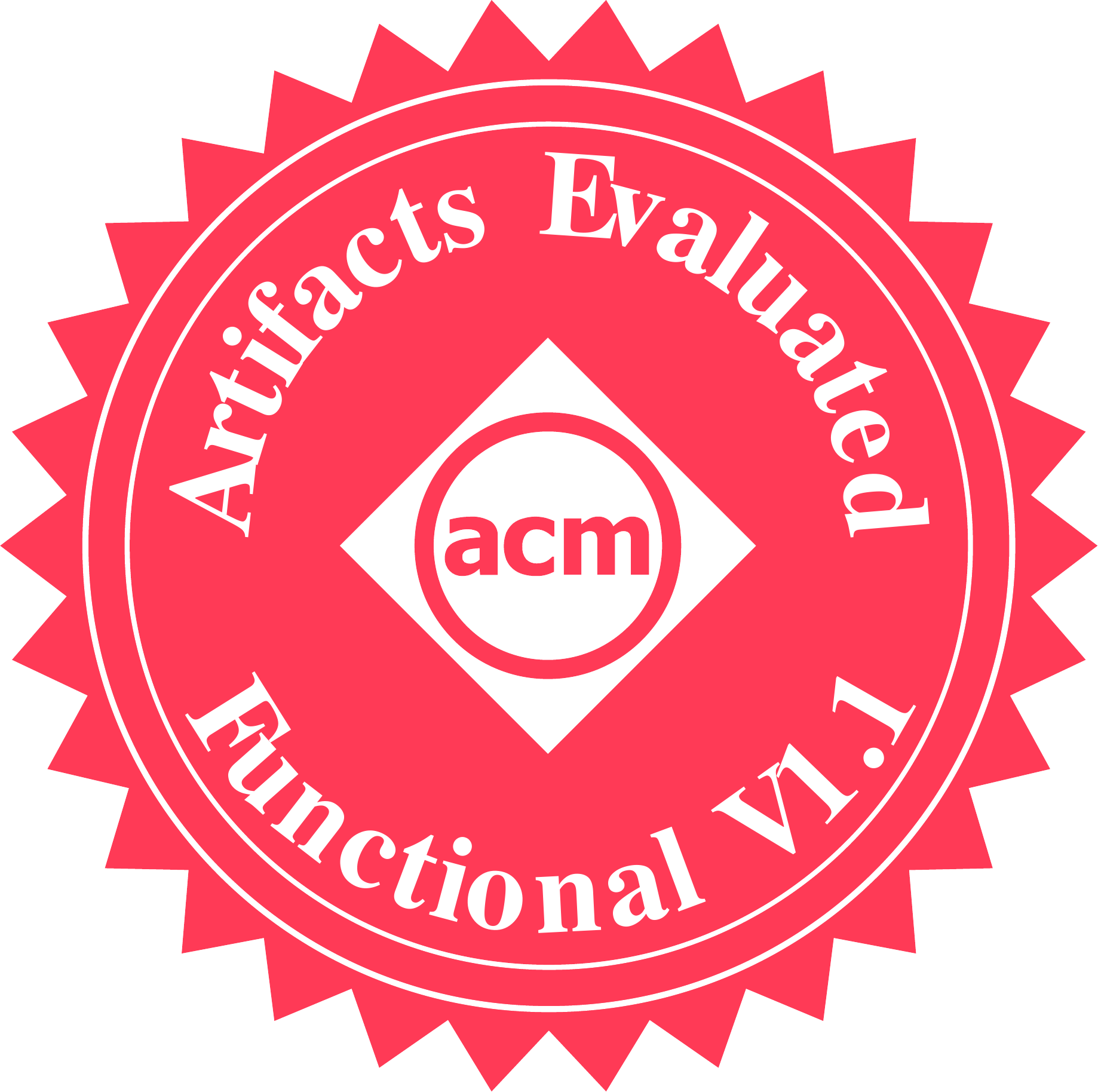}
}};
\end{tikzpicture}
}}
\BgThispage

\section{Introduction}\label{sec:Intro}

\begin{figure*}[b]
    \centering
    \includegraphics[width=0.85\textwidth]{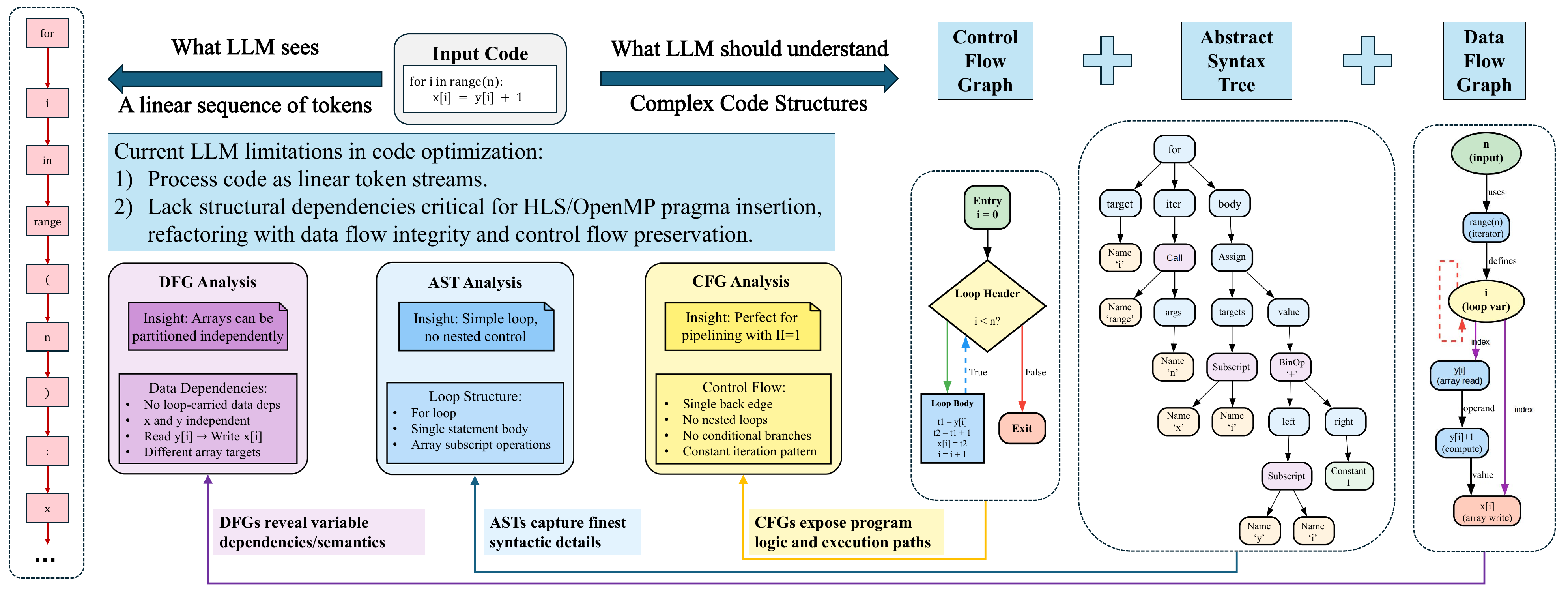}
    \caption{What LLM Fails to See in the Code.}
    \label{fig:motivation}
    \Description{}
\end{figure*}

High-Level Synthesis (HLS) raises the abstraction level of hardware design from register-transfer logic to C/C++, but the quality of the resulting hardware is almost entirely determined by a small set of optimization directives, pragmas. Whether a loop is pipelined, how aggressively it is unrolled, and how its arrays are partitioned across memory banks determine the throughput, latency, and resource footprint of the synthesized accelerator, often by an order of magnitude. A single misplaced \texttt{ARRAY\_PARTITION} can leave a pipelined loop starved for memory bandwidth; a single \texttt{PIPELINE} on a loop with a carried dependence can inflate the initiation interval from one cycle to dozens. 
Currently,  pragma insertion remains one of the most labor-intensive steps in the HLS design flow, and automating it is a central problem for machine learning in CAD~\cite{cong2011hls, RALAD}.

Automating this reasoning has been a central goal of machine learning for CAD. The dominant approach treats pragma selection as design-space exploration (DSE): enumerate candidate configurations, estimate the quality of results (QoR) of each with a fast surrogate, and search. AutoDSE~\cite{autodse} formalized the search; GNN-DSE~\cite{sohrabizadeh2022gnndse}, HARP~\cite{sohrabizadeh2023harp}, and IronMan~\cite{Ironman} replaced the surrogate with a graph neural network that reads the program graph and predicts latency and resource usage with remarkable fidelity. These methods, although structure-aware, score or rank a fixed candidate configuration rather than generate new directives and require an external design‑space‑exploration loop.

Large language models (LLMs) excel at generic code generation, yet when they read a program, they see only a linear token sequence.
What an LLM should see (as illustrated in Figure~\ref{fig:motivation}) for reliable HLS optimization is the complex, hierarchical structure of the code $-$ its Abstract Syntax Tree (AST), Control Flow Graph (CFG), and Data Flow Graph (DFG) $-$ which traditional compilers use to guarantee correctness and to reason about performance.
Because this structural information is absent, LLMs frequently hallucinate illegal or sub-optimal \emph{pragmas} that only fail during synthesis.
%

In this paper, we present a novel framework that injects \emph{three} hierarchical program graphs $-$ AST, CFG, and DFG $-$ into a pretrained code LLM.
The graphs are encoded by the three parallel GNN streams, compressed into a compact token set, and merged with the LLM via a \emph{gated cross-attention} layer that lets the model fall back to its generic coding knowledge when the structural signal is weak.
By training the graph encoders and the LLM jointly but on separate parameter subsets, our method learns program structure and pragma intent as interacting yet independent modalities. Our contributions include:
\begin{itemize}[topsep=1pt]
  \item First joint integration of AST, CFG, and DFG of a code into an LLM for HLS pragma generation with a gated cross-attention mechanism that preserves the LLM’s baseline coding ability while exploiting structural information.
  \item Decoupled, jointly-trained architecture that treats program structure and optimization directives as separate but interacting modalities.
  \item It synthesizes 3.5× as many kernels as its Llama-base ancestor (26.9\% vs. 7.7\%), and on the kernels where it does succeed, it produces designs that are 2.31× faster (geomean) than GPT-5-mini's.
  \item On the agentic flow, it outperforms other baselines when optimizing complex code and achieves the average normalized improvement across a 26-kernel suite to 26.4\%.
\end{itemize}

The remainder of the paper is organized as follows: Section~\ref{sec:Background} discusses current state-of-the-art work,  Section~\ref{sec:Design} describes our structure-augmented LLM architecture, Section~\ref{sec:Experiments} details the training methodology and presents the experimental results, and Section~\ref{sec:conclusion} concludes with a discussion of future work.

\section{Background}\label{sec:Background}
Existing research falls into two complementary strands: (a) learning-based surrogate models that predict Quality-of-Result (QoR) for a given pragma configuration, and (b) large language models (LLMs) that edit or generate source code but typically ignore the underlying program structure.
We briefly review each strand and highlight the gap our work fills.

\subsection{Learning-Based HLS Design-Space Exploration}

To avoid invoking the synthesizer for every candidate, many works train surrogate models that predict QoR from a program representation. AutoDSE~\cite{autodse} framed the problem as a guided search over the pragma space. GNN-DSE~\cite{sohrabizadeh2022gnndse} builds a graph from the kernel's LLVM IR and trains a GNN to regress latency and resource usage; a downstream search loop then queries the surrogate instead of the synthesizer.
HARP~\cite{sohrabizadeh2023harp} refines this idea with a hierarchical graph that isolates program semantics from pragma nodes, thereby improving transfer across kernels. IronMan~\cite{Ironman} coupled a GNN predictor with reinforcement learning to drive the search itself. 
Although accurate and structurally faithful, these models are fundamentally discriminative: they evaluate only configurations proposed by an external search procedure and never emit an optimized source.

Our structure-augmented model complements these surrogates by directly producing a pragma-annotated kernel; a surrogate such as HARP could subsequently be used as a verifier or reward signal.

\subsection{LLMs for Code and for Hardware Design}


Instruction-tuned code LLMs have made code editing and generation accessible via natural language. Code Llama~\cite{codellama} and
StarCoder~\cite{li2023starcodersourceyou} are strong open baselines, and instruction-tuning corpora such as InstructCoder~\cite{li2024instructcoderinstructiontuninglarge} adapt general-purpose LLMs to edit existing code rather than generate it from scratch.
In the hardware domain, ChipNeMo~\cite{liu2024chipnemodomainadaptedllmschip} adapts LLMs to EDA tasks through domain pretraining, and several systems apply LLMs to RTL and HLS generation~\cite{thakur2023verigenlargelanguagemodel,lu2023rtllmopensourcebenchmarkdesign, gpt4aigchip}.
All of these systems consume code as a flat token sequence.
While sufficient for surface-level edits, this representation provides no explicit view of control- and data-flow — information that is essential for deciding whether a pragma is legal and beneficial.

\subsection{Structure-Aware Models for Code}
\label{sec:struct-code}
The intuition that programs are graphs, not just strings, predates LLMs. Code2vec~\cite{alon2018code2veclearningdistributedrepresentations} learned representations from AST paths; GraphCodeBERT~\cite{guo2021graphcodebertpretrainingcoderepresentations} added data-flow edges to a transformer encoder during pretraining; and StructCoder~\cite{structcoder} conditioned a sequence-to-sequence model on AST and DFG signals. These works establish that structure helps, but they share three properties that limit their use for HLS. First, they inject structure on the encoder side, at pretraining time, which couples the structural component to a particular base model and an expensive training run. Second, they use a single graph view (usually the DFG, whereas a compiler reasons over syntax, control, and data flow together). Third, they target understanding and short edits, not the long, fragile directive blocks HLS requires.

\section{Structure-augmented LLMs}
\label{sec:Design}
\subsection{Overview}
\label{sec:gaco-overview}
\begin{figure}[b]
    \vspace{-3ex}
    \centering
    \includegraphics[width=1.05\linewidth]{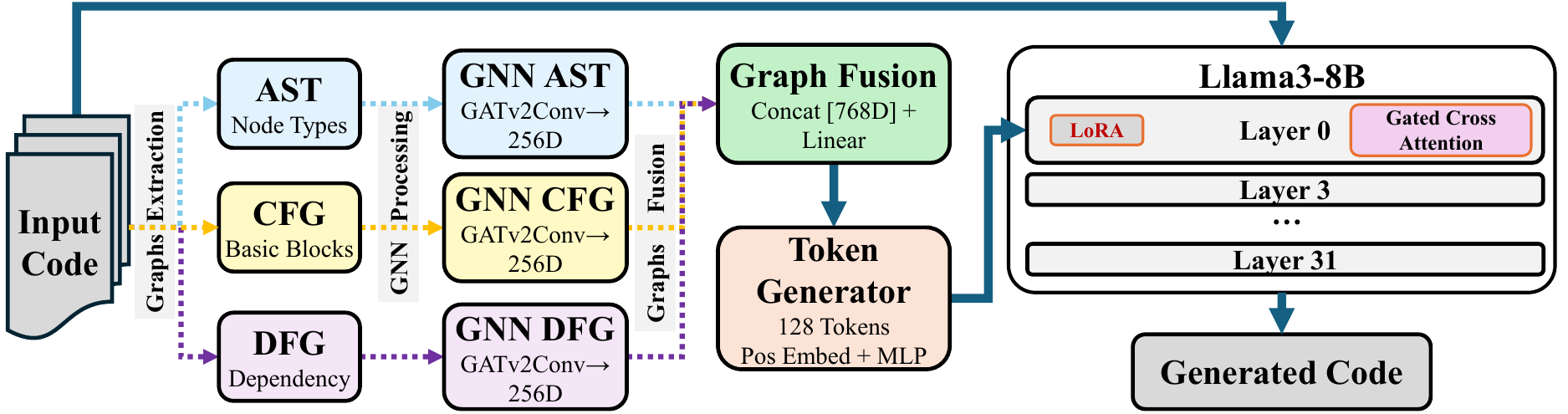}
    \caption{Structure-augmented LLM with Layer-specific Graph Injection}
    \label{fig:arch}
    \Description{}
\end{figure}

As shown in Figure~\ref{fig:arch}, \sysname{} (Pragma Reasoning via Integrated Structural Models) augments a pretrained, frozen code LLM with an explicit structural view of its input. 
Given a source kernel $s$ and an optimization instruction, our structure-augmented model proceeds in four stages.
First, it extracts three complementary graphs (AST, CFG, DFG) from $s$.
Second, each graph is processed by its own GATv2 encoder~\cite{brody2022gatv2}, yielding a fixed-length embedding $z_g \in \mathbb{R}^{256}$ (Section~\ref{sec:gaco-gnn}).
Third, the three embeddings are concatenated and projected into a bank of $k = 128$ graph tokens $\mathbf{T} \in \mathbb{R}^{k \times d}$ that reside in the LLM's embedding space (Section~\ref{sec:gaco-fusion}).
Finally, the LLM’s text hidden states attend to $\mathbf{T}$ through a gated cross-attention module; the resulting enriched states are fed to the remaining transformer layers for decoding (Section~\ref{sec:gaco-xattn}).
The complete data path is shown in Figure~\ref{fig:arch}.
We instantiate the base model as
Llama-3.1-8B-Instruct with hidden size $d{=}4096$; all of its weights are frozen, and only the graph encoders, the projector, the cross-attention and gating parameters, and low-rank adapters (LoRA) are trained.

Two principles shape the design, and each anticipates a failure mode we wanted to avoid: \textbf{Decouple structure from directives.} The text stream carries the instruction and the kernel as the model learned to read them; the graph stream carries an orthogonal, compiler-grade view of the same kernel, keeping the two separate until they are fused, rather than serializing the graph into the prompt. Let each be learned with its own parameters, and let the model learn their interaction explicitly, rather than forcing the structure through the same tokenizer that already struggles with it.
\textbf{Preserve the pretrained model.} A code LLM's value lies in weights that took enormous compute to train. Any structural addition that overwrites them trades one competence for another. \sysname{} therefore freezes the base model and admits structural signal only through gated residual paths (Section~\ref{sec:gaco-xattn}), so that in the limit of an unhelpful gate, the model reduces exactly to the original LLM.

 
\subsection{Graph Construction}\label{sec:gaco-graphs}
From the source kernel, we extract three different graphs, each exposing a distinct facet of the program, illustrated for a simple loop in Figure~\ref{fig:motivation}.
The abstract syntax tree (AST) captures the hierarchical syntax, such as loop nests, subscript expressions, and operator structure. It shows that for the loop \texttt{x[i] = y[i] + 1}, the body is a single statement over array subscripts. 
The control‑flow graph (CFG) records execution paths, including basic blocks, back edges, and branch conditions, making loop trip counts and nesting readily visible.
Finally, the data‑flow graph (DFG) represents definition‑use relationships, thereby exposing loop‑carried dependencies and array read/write patterns. In this case, it exposes that \texttt{x} and \texttt{y} are independent with no loop-carried dependence, which is precisely the fact that licenses partitioning the arrays and pipelining at II${=}1$.
These are the inferences an HLS expert makes at a glance, and that a
token-streaming model has no direct way to make.

For Python kernels, the AST is obtained from the native \texttt{ast} module, and the CFG and DFG are constructed by a single traversal of that tree.
The C/C++ extraction pipeline is described in Section~\ref{sec:gaco-transfer}.
Each node receives a feature vector formed by concatenating a one-hot encoding of its node type, a hash of its surface token, and its degree.
This yields a per-graph node-feature matrix $X_g \in \mathbb{R}^{n_g \times f}$ and an edge-index $E_g$ for $g \in \{\textsc{ast}, \textsc{cfg}, \textsc{dfg}\}$.
 
\subsection{Per-Graph GNN Encoders}
\label{sec:gaco-gnn}
Each graph is encoded independently by a stack of GATv2~\cite{brody2022gatv2} convolutions. GATv2 is chosen over the original GAT because its dynamic attention can express edge-conditioned importance, which matters when, for example, a single back edge in the CFG must dominate the loop's representation. A stack uses multi-head attention (four heads) in the first layer, projects to a single head in the second, applies graph normalization, and pools the node embeddings into a graph-level vector $z_g \in \mathbb{R}^{256}$:
\begin{equation}
  z_g = \operatorname{Pool}\!\big(\operatorname{GraphNorm}(\operatorname{GATv2}(X_g, E_g))\big),
  \qquad g \in \{\textsc{ast}, \textsc{cfg}, \textsc{dfg}\}.
  \label{eq:gnn}
\end{equation}

Encoding the three graphs in parallel with separate parameters prevents cross-modal contamination and lets the subsequent fusion module learn explicit interactions among the modalities.
 
\subsection{Graph Fusion and Token Generation}
\label{sec:gaco-fusion}
The three graph vectors are concatenated into a single structural descriptor $z = [\,z_{\textsc{ast}}; z_{\textsc{cfg}}; z_{\textsc{dfg}}\,] \in \mathbb{R}^{768}$,
which a learned \emph{graph token generator} expands into a bank of $k = 128$ tokens in the LLM embedding space. Concretely, a linear projection followed by a positional embedding and an MLP maps $z$ to
$G \in \mathbb{R}^{k \times d}$:
\begin{equation}
  G = \operatorname{MLP}\!\big(\operatorname{Proj}(z) + P\big),
  \qquad G \in \mathbb{R}^{128 \times 4096},
  \label{eq:tokens}
\end{equation}
where $\mathbf{P}$ are learned positional embeddings over the $k$ token slots. 
Expanding a single 768-d descriptor into $k$ tokens (rather than one) gives the cross-attention layer multiple structural ``slots'' to attend to, which we found necessary for the model to route different aspects of structure to different parts of the generated kernel.
 
\subsection{Gated Cross-Attention Injection}
\label{sec:gaco-xattn}

We inject the graph tokens at layer~0. By front-loading all structural information, we guarantee that every subsequent layer receives an enriched representation, while keeping the pre-trained weights intact through a lightweight gating mechanism that controls how much of the new token the model can use.

Let $H \in \mathbb{R}^{m \times d}$ be the text hidden states over the $m$ input tokens and $G \in \mathbb{R}^{k \times d}$ the graph tokens. The module first computes a standard cross-attention with the text states as queries and the graph tokens as keys and values,
\begin{equation}
  A = \operatorname{softmax}\!\left(\frac{(H W_Q)(G W_K)^\top}{\sqrt{d_k}}\right) G W_V,
  \label{eq:xattn}
\end{equation}
and then blends the attended structural features $A$ with the original states $H$ under a learned, element-wise gate:
\begin{align}
  \gamma &= \sigma\!\big(\operatorname{MLP}([\,H; A\,])\big), \label{eq:gate}\\
  H' &= \gamma \odot A + (1 - \gamma) \odot H. \label{eq:blend}
\end{align}
Here $\sigma$ is the logistic function and $\odot$ is the Hadamard product, so $\gamma \in (0,1)$ controls, per position and per channel, how much structural signal is admitted. When $\gamma \to 0$ the layer reduces to the identity and the model behaves exactly as the pretrained LLM; this is what allows our structure-augmented model to add structure without overwriting the base model's coding ability and what keeps training stable, since the network can ignore an unhelpful graph signal rather than being forced to use it. The enriched states $H'$ are then passed to the remaining transformer layers and decoded.

\subsection{Model Fine-tuning}
\label{sec:gaco-train}
\sysname{} is trained end-to-end with the base LLM frozen. We attach LoRA~\cite{hu2021loralowrankadaptationlarge} adapters to the transformer layers and update only the adapters, the three GNN encoders, the token generator, and the cross-attention and gating parameters. Freezing the base both slashes the trainable parameter count and guards against catastrophic forgetting of the pretrained coding knowledge that \sysname{} is built to preserve. Because the components converge at different rates, for example, a randomly initialized GNN learns faster and noisier than a LoRA adapter on a converged backbone. Thus, we give each component its own learning-rate schedule, pacing the graph encoders and cross-attention heads to track the LoRA updates so that the structural and textual pathways advance together rather than one destabilizing the other. We fine-tune on InstructCoder~\cite{li2024instructcoderinstructiontuninglarge}, a corpus of over 114k instruction--input--output Python triples for code editing, optimization, and refactoring, which matches \sysname{}'s edit-an-existing-kernel usage. There is no additional fine-tuning on the C/C++ HLS pragma
corpus. 
 
\subsection{Cross-Language Structural Transfer}
\label{sec:gaco-transfer}
Because the graph encoders were trained on Python, applying our structure-augmented model to C/C++ HLS kernels requires bridging two graph ``languages.'' 
We first strip macros, pragmas, and comments from the C source, rebuild the AST, CFG, and DFG over the cleaned code, and map C node types onto the Python node-type vocabulary so the encoders receive embeddings drawn from their training distribution. 
The mapping is imperfect because C exposes constructs with no Python analogue, notably explicit variable declarations and initializations, and these mismatches account for residual error quantified in Section~\ref{sec:eval}. 
In all experiments, we retain the bare (pragma-free) kernel as the structural source, so the graph stream encodes the program to be 
optimized rather than any particular pragma choice, aligning with our model's goal of conditioning generation on input structure.

\section{Experiments and Results}\label{sec:Experiments}
\label{sec:eval}
Our evaluation answers two questions. 
First, does the structural signal enable the model to generate a synthesizable, optimized pragma block in a single shot where the baseline model fails (Section~\ref{sec:one-shot})?
Second, does this advantage persist (or change character) when the model iterates and evolves within an agentic flow (Section~\ref{sec:agentic})?

\subsection{Zero-shot evaluation}
\label{sec:one-shot}

\subsubsection{Setup}
\label{sec:eval-setup}
We compare \sysname{} against three baselines that span open and closed, small and large: \textbf{Llama-3.1-8B}~\cite{llama3} (the base of \sysname{}); 
\textbf{gpt-oss-20b}~\cite{GPTOSS}, a 20B-parameter open model served locally; and 
\textbf{GPT-5-mini}~\cite{GPT5mini}, a frontier proprietary model served via the OpenAI API. Local models are evaluated on a single NVIDIA 4090 GPU. 
Each model is asked to emit a pragma block for a kernel; the generated source then runs through the HLS toolchain.

We use the HLS-Eval suite~\cite{HlsEval} of 106 kernels drawn from five benchmark families: PolyBench~\cite{polybench}, MachSuite~\cite{MachSuite}, CHStone~\cite{CHStone}, Rosetta~\cite{Rosetta}, and ForgeBench~\cite{wanna2025forgebenchmachinelearningbenchmark}. The four models attempt slightly different subsets (some kernels produce malformed generations under one model and not another), so for an apples-to-apples comparison, we restrict the pass-rate and latency analyses to the 78-kernel intersection that every model attempted. We score three outcomes per generation: \textbf{Compile} — the emitted C compiles and passes a software test bench; \textbf{Synth} — the design synthesizes to RTL under the HLS tool; and report \textbf{Latency} for synthesized designs.


\subsubsection{Python Benchmark}

Before turning to HLS, we verify that the structural signal does not degrade general code-editing ability. We evaluate \sysname{} and the Llama-8B~\cite{llama3} baseline on HumanEval~\cite{chen2021evaluatinglargelanguagemodels}, 164 hand-written Python tasks with reference tests. We measure syntactic validity (via \texttt{ast.parse}) and Pass@1 under two output-handling strategies: \emph{simple concatenation}, which appends the raw generation, and \emph{clean extraction}, which truncates at natural stopping markers. The distinction matters: it isolates whether gains come from reasoning or from generation discipline.

As shown in Table~\ref{tab:baseline}, under clean extraction, the two models are statistically tied (97.6\% vs.\ 98.8\% Pass@1). Under simple concatenation, the Llama baseline collapses to 55.5\% Pass@1 because 93.9\% of its outputs over-generate spurious test code; \sysname{}{} holds at 97.0\%, with over-generation in only 1.8\% of outputs. 
The 41.5-percentage-point gap (about a 75\% relative improvement) is therefore a generation-discipline effect (structural grounding teaches the model where the program ends) rather than a reasoning lift on easy Python. 
The inference-time cost is real but bounded: \sysname{} runs about $3\times$ slower per query than the base (20.45\,s vs.\ 6.75\,s in our setup).

\begin{table}[t]
\centering
\caption{HumanEval Llama3-8B vs \sysname{}}
\label{tab:baseline}
\vspace{-2ex}
\small
\begin{tabular}{|l|cc|cc|c|}
\toprule
 & \multicolumn{2}{c|}{\textbf{Syntax Validity}} & \multicolumn{2}{c|}{\textbf{PASS$@$1}} & \multicolumn{1}{c|}{\textbf{Avg. Time}} \\
\textbf{Model} & Simple & Clean & Simple & Clean & (s)  \\
\midrule
Llama3-8B & 63.4\% & 97.6\% & 55.5\% & 97.6\% & 6.75 \\
\sysname{} & 98.2\% & 98.8\% & 97.0\% & 98.8\% & 20.45  \\
\bottomrule
\end{tabular}
\vspace{-2ex}
\end{table}

\begin{figure}[b]
    \centering
    \includegraphics[width=\linewidth]{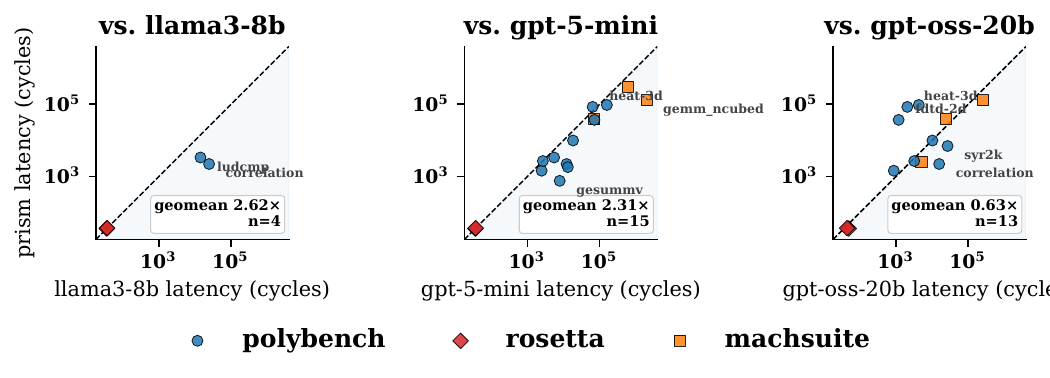}
    \caption{Latency Speedup of \sysname{}}
    \label{fig:speedup_scatter}
\end{figure}

\begin{table*}[t]
  \centering
  \caption{Zero-shot pragma-insertion pass rates on the intersection of 78 kernels that every model attempted (apples-to-apples). \textbf{Synth}: the generated kernel synthesizes to RTL under the HLS tool. \textbf{Compile} (in italics, denoted cmp): the generated C compiles and passes the software test bench.}
  \label{tab:pass-rates-intersection}
  \setlength{\tabcolsep}{4pt}
  \small
  \begin{tabular}{lccccccc}
    \toprule
    \textbf{Model} & \textbf{Polybench(24)} & \textbf{Machsuite(16)} & \textbf{Chstone(18)} & \textbf{Rosetta(8)} & \textbf{Forgebench(12)} & \textbf{Overall(78)} \\
    \midrule
    \textbf{\sysname{}}  & 37.5\% (9/24) & 43.8\% (7/16) & 0.0\% (0/18) & 62.5\% (5/8) & 0.0\% (0/12) & 26.9\% (21/78) \\
     & \textit{(92\% cmp)} & \textit{(62\% cmp)} & \textit{(33\% cmp)} & \textit{(62\% cmp)} & \textit{(0\% cmp)} & \textit{(55\% cmp)} \\[2pt]
    Llama3-8B  & 12.5\% (3/24) & 6.2\% (1/16) & 0.0\% (0/18) & 25.0\% (2/8) & 0.0\% (0/12) & 7.7\% (6/78) \\
     & \textit{(75\% cmp)} & \textit{(44\% cmp)} & \textit{(0\% cmp)} & \textit{(25\% cmp)} & \textit{(0\% cmp)} & \textit{(35\% cmp)} \\[2pt]
    gpt-oss-20B  & 79.2\% (19/24) & 31.2\% (5/16) & 16.7\% (3/18) & 87.5\% (7/8) & 0.0\% (0/12) & 43.6\% (34/78) \\
     & \textit{(83\% cmp)} & \textit{(75\% cmp)} & \textit{(50\% cmp)} & \textit{(88\% cmp)} & \textit{(42\% cmp)} & \textit{(68\% cmp)} \\[2pt]
    gpt-5-mini  &91.7\% (22/24) & 62.5\% (10/16) & 11.1\% (2/18) & 75.0\% (6/8) & 8.3\% (1/12) & 52.6\% (41/78) \\
     & \textit{(100\% cmp)} & \textit{(88\% cmp)} & \textit{(56\% cmp)} & \textit{(88\% cmp)} & \textit{(17\% cmp)} & \textit{(73\% cmp)} \\[2pt]
    \bottomrule
  \end{tabular}
\end{table*}

\subsubsection{HLS Synthesis Pass Rates}
Table~\ref{tab:pass-rates-intersection} reports Synth and Compile rates over the 78-kernel intersection across all four models. We found 
\textbf{1) Structural augmentation lifts a small base model substantially.}
Going from Llama-3.1-8B (7.7\% synth) to \sysname{} (26.9\% synth) is a $3.5\times$ improvement on the kernels both models attempted with structural augmentation (Section ~\ref{sec:Design}). The gain is consistent across PolyBench~\cite{polybench}, MachSuite~\cite{MachSuite}, and Rosetta~\cite{Rosetta}. 
\textbf{2) Frontier and larger open baselines retain a coverage advantage.} GPT-5-mini reaches 52.6\% and gpt-oss-20b reaches 43.6\% on the same intersection — both higher than \sysname{}'s 26.9\%. We do not claim \sysname{} beats them on raw synth coverage; the contribution is elsewhere (Section~\ref{sec:eval-latency}). 
What this comparison does establish is that an 8B parameter, locally-served, structurally-grounded model closes a meaningful fraction of the gap to a far larger proprietary model without relying on its scale (roughly 2.5$\times$ fewer parameters).
\textbf{3) Per-family behavior isolates where structure pays.} \sysname{} outperforms gpt-oss-20b on MachSuite (43.8\% vs.\ 31.2\%), where loop and array patterns are rich enough for the structural signal to exploit. It trails on PolyBench, whose regular nested loops are forgiving enough that scale alone suffices, and collapses to 0\% on CHStone, which is a benchmark heavy in C-specific constructs (struct definitions, typedefs, explicit pointer arithmetic) for which the Python-trained graph encoders have no analogue. 
This is exactly the failure mode predicted by our cross-language transfer scheme (Section~\ref{sec:gaco-transfer}). The fact that GPT-5-mini reaches 11.1\% on CHStone isolates the residual gap to precisely the construct class the node-type mapping does not cover, and points to native C/C++ graph encoders as the clearest path to closing it (Section~\ref{sec:conclusion}). ForgeBench remains near zero for all 4 models and yields only one synthesizable design, even for GPT-5-mini, which needs further investigations.
 
\subsubsection{HLS Latency Quality}
\label{sec:eval-latency}
Synth rate alone does not characterize the quality of a generated design.
A pragma block that synthesizes may produce RTL that meets the toolchain's correctness gate but barely improves on the no-pragma baseline. To compare quality directly, Figure~\ref{fig:speedup_scatter} plots, for each baseline, \sysname{}'s synthesis cycle count against the baseline's on the kernels where both synthesized. Both axes are log-scale; the dashed line is $y=x$; the shaded region below the line is where \sysname{} produces a faster design.
\textbf{Against the Llama-base ancestor, \sysname{} is uniformly faster}~(geomean $2.62\times$, $n{=}4$). Every shared synthesized kernel sits below the diagonal. Structural augmentation does not merely raise the chance of a legal design; it raises the quality of the legal designs it produces. \textbf{Against GPT-5-mini, \sysname{} is faster on the geomean despite the frontier model's coverage advantage} (geomean $2.31\times$, median $1.80\times$, $n{=}15$). The largest single-kernel wins are \texttt{machsuite\_\_gemm\_ncubed} ($15.2\times$), \texttt{polybench\_\_gesummv} ($10.3\times$), \texttt{polybench\_\_covariance} ($7.3\times$), and \texttt{polybench\_\_correlation} ($5.5\times$). The single notable loss is \texttt{polybench\_\_heat-3d} ($0.76\times$). The pattern is consistent: on kernels with the loop-and-array structure the graph stream can exploit, \sysname{}'s smaller, structurally-grounded model finds optimizations that a
roughly order-of-magnitude larger general model misses.
\textbf{Against gpt-oss-20b, results are mixed} (geomean $0.63\times$, median $1.19\times$, $n{=}13$). \sysname{} wins on the median kernel but loses heavily on three PolyBench stencils (\texttt{heat-3d} ($0.02\times$), \texttt{jacobi-2d} ($0.04\times$), and \texttt{fdtd-2d} ($0.03\times$)) for which gpt-oss-20b reports cycle counts an order of magnitude below what the iteration counts alone would predict. These designs pass synthesis and the test bench, but the reported latencies suggest an aggressive scheduling that may compress the iteration space rather than execute it; we report the geomean as-is for transparency and flag this caveat for future investigation.

\begin{table*}[t]
  \centering
  \caption{Agentic adaptation over 26 kernels with 3 iterations.
    \textbf{Synth}/\textbf{Correct} are the fractions of runs that synthesize /
     pass functional verification. \textbf{Avg. Improvement} is the average of normalized improvement over all kernels. \textbf{Total Tokens} is total tokens consumed. \textbf{Correct/M Tok} is the amount of correct design per million tokens. }
  \label{tab:agentic-results}
  \small
  \setlength{\tabcolsep}{6pt}
  \vspace{-1ex}
  \begin{tabular}{cccccccc}
    \toprule
    \textbf{Config} & \textbf{Planner} & \textbf{Codegen}
                 & \textbf{Synth} & \textbf{Correct} & \textbf{Avg. Improvement} & \textbf{Total Tokens} & \textbf{Correct/M Tok}\\
    \midrule
   A & Llama-base         & \sysname{}         & \SI{11.7}{\percent} & \SI{11.1}{\percent} & 9.5\% & 622{,}353 & 5.36 \\
    B & \textsc{lora-text} & \textsc{lora-text} & \SI{13.0}{\percent} & \SI{11.3}{\percent} & 14.2\% & 407{,}443 & 8.32 \\
    C & \textsc{lora-text} & \sysname{}         & \SI{19.9}{\percent} & \SI{19.9}{\percent} & 14.1\% & 631{,}691 & \textbf{9.45}\\
    D & GPT-5-mini         & \sysname{}         & \SI{27.8}{\percent} & \SI{27.8}{\percent} & 26.4\% & 1{,}442{,}718 & 5.77 \\

E & \sysname{}         & \sysname{}         & \textbf{\SI{32.2}{\percent}} & \textbf{\SI{32.2}{\percent}} & 0\% & 437{,}377 & 22.1 \\
    \bottomrule
  \end{tabular}
\end{table*}
\subsection{Iterative Refinement via Agentic Adaptation}
\label{sec:agentic}
The zero-shot results establish that \sysname{}{} produces high-quality designs when it succeeds but at lower coverage than larger baselines (Section~\ref{sec:one-shot}). In practice, HLS optimization is iterative: a designer (or an agent) proposes a pragma plan, reads the synthesizer's feedback, and revises. We now ask how structural conditioning behaves inside that loop. We report two complementary experiments. Section~\ref{sec:agentic-convergence} runs each model as both planner and codegen in a closed loop on three kernels of increasing structural complexity, exposing per-kernel convergence dynamics. Section~\ref{sec:agentic-aggregate} factorizes the loop into a $5{\times}26$ planner-codegen study that decouples the two roles across 26 kernels and isolates where the structural signal pays.


\subsubsection{Setup}
\label{sec:agentic-setup}
The agentic loop alternates between a \emph{planner} role that proposes a pragma strategy in natural language and a \emph{codegen} role that emits the pragma block. After each iteration, the design is synthesized; the planner receives the synthesis log and achieved latency in the next round. We run ten refinement iterations per kernel in Section~\ref{sec:agentic-convergence} and three iterations in Section~\ref{sec:agentic-aggregate}. We report two metrics: \textbf{improvement\%} $= 100 \times (1 - \text{best latency} / \text{baseline latency})$, normalized per kernel; and \textbf{synth pass rate}, the fraction of kernels for which the loop produces at least one synthesizable design. The five configurations in Table~\ref{tab:agentic-results} factorize structural conditioning across the planner and codegen, with \textsc{lora-text} denoting the same checkpoint as \sysname{} but served with the graph encoder disabled.

 \begin{figure}
     \centering
     \includegraphics[width=\linewidth]{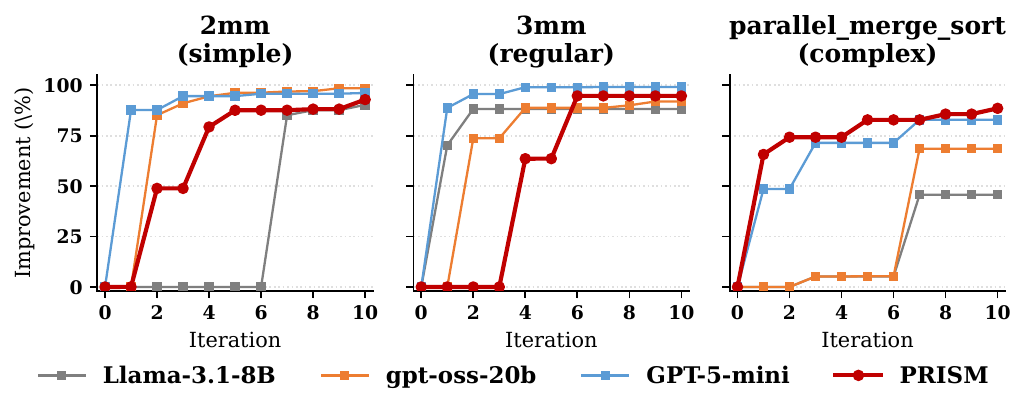}
     \caption{Agentic curve - Improvement vs iterations}
     \label{fig:agentic-curves}
 \end{figure}

        

\subsubsection{Convergence depends on structural complexity}
\label{sec:agentic-convergence}
We first run single-model loops in which each model serves as codegen with gpt-5-mini as planner on three kernels of increasing structural complexity: \texttt{2mm} (regular triply-nested matrix chain), \texttt{3mm} (slightly larger but similarly regular), and \texttt{parallel\_merge\_sort} (irregular control flow with recursive splitting and dynamic memory access patterns). Figure~\ref{fig:agentic-curves} plots improvement over the baseline against iteration count.
 
The model ordering reshuffles with kernel complexity. On \texttt{2mm}, GPT-5-mini and gpt-oss-20b reach 50\% improvement within one to two iterations while \sysname{} lags to iteration~4 and finishes mid-pack (92.9\% vs.\ 98.6\% for the leader). On \texttt{3mm} the ordering is similar. On \texttt{parallel\_merge\_sort}, the picture inverts. \sysname{} is the only model to improve on iteration~1 (reaching 65.7\% on its first pragma proposal) reaches the 50\% threshold three iterations earlier than GPT-5-mini and six earlier than gpt-oss-20b, and finishes at the highest final improvement (88.6\% vs.\ 82.9\% for GPT-5-mini, 68.6\% for gpt-oss-20b, and 45.7\% for the Llama-3.1-8B baseline). The structural priors deliver their advantage exactly where the underlying task most rewards structural reasoning, mirroring the per-family pattern observed in zero-shot synthesis.
 
We are careful not to overclaim: \sysname{} is not the strongest single-model agentic loop in absolute terms. It is the strongest when the kernel genuinely benefits from structural reasoning, and its iteration efficiency scales with that structural payoff. This dynamic motivates the factorized study below, which asks whether structural conditioning is best deployed as a planner or as a codegen step.


 
\subsubsection{Where structural conditioning belongs}
\label{sec:agentic-aggregate}
Table~\ref{tab:agentic-results} aggregates over all 26 kernels, with three findings that stand out.

\textbf{Finding 1: structural codegen lifts legality without distorting the planner's strategies, and it scales with planner strength}. Configurations~B and~C share the \textsc{lora-text} planner and differ only in the codegen. Swapping the codegen from \textsc{lora-text} to \sysname{} raises the correct rate from \SI{11.3}{\percent} to \SI{19.9}{\percent} (a \SI{8.6}{percentage-point} gain) and correct per million tokens from 8.32 to 9.45 while leaving the average improvement essentially unchanged (1\% difference). The structural signal makes the codegen better at producing legal, synthesizable pragmas while maintaining similar optimizations the planner asks for. From A to C to D, correctness and improvement steadily increase when the planner gets stronger, which suggests \sysname's scalability.  
 
\textbf{Finding 2: structural planning sacrifices optimization for legality.} Configurations~D and~E share the \sysname{} codegen and differ only in the planner. 
Although the \sysname{} planner (E) generates a larger share of correct designs than the GPT-5-mini planner (D) (32.2\% vs. 27.8\%), it shows no average improvement, whereas D achieves a 26.4\% improvement. 
This is expected: \sysname{} has been fine-tuned as a structure-conditioned model, so its performance degrades in the purely text-based planning stage.
The structurally grounded planner converges quickly to a single safe plan that satisfies the synthesizer but fails to push the QoR on any kernel beyond what the baseline already achieves.
Plan diversity, which natural-language reasoning supplies and structural conditioning suppresses, is what enables deep optimization in the iterative loop. We also found GPT-5-mini as the best Planner in all cross-validation runs. 
 
\textbf{Finding 3: structural conditioning is a role-specific tool.} Read together, the B\,$\to$\,C and D\,$\to$\,E ablations identify where the structural signal belongs. The codegen role rewards legality, and structural conditioning supplies that legality without cost. The planner role rewards strategy diversity, and structural conditioning withdraws it. The best configuration in the suite is therefore~D: a natural-language planner paired with the \sysname{} codegen. We did not anticipate this result at the start of the project, where the symmetric design (\sysname{} on both sides) was the natural default, and we believe the asymmetric lesson it teaches is the most transferable contribution of our agentic study. Any system that adds a structural adapter to an agentic LLM loop should ask, for each role, whether that role rewards exploration or rewards constraint, and place the adapter accordingly.
 
\noindent\textbf{An efficiency note.} Token totals in Table~\ref{tab:agentic-results} do not track quality of result monotonically. The most successful configuration (D) is also the most expensive (\SI{1.44}{M} tokens), and the cheapest \sysname{}-codegen configuration (E) produces no improvement at all. Tokens are the cost of exploration in this setting, and exploration is what differentiates D from E. A practitioner choosing between D and the more economical C trades roughly $2.3\times$ the token budget for roughly $1.9\times$ the average improvement and $1.4\times$ the correct rate $-$ a steep but justifiable bill for kernels where QoR matters.

\section{Conclusion}
\label{sec:conclusion}

We presented \sysname{}, a framework that augments a pretrained, frozen code-LLM with explicit structural reasoning for HLS-pragma optimization.
By encoding the kernel's AST, CFG, and DFG with three parallel GNN streams and injecting them at a specific transformer layer through gated cross-attention, \sysname{} bridges the long-standing divide between structure-aware models that cannot generate and generative models that are structure-blind. 
In zero-shot evaluation, it synthesizes 3.5× as many kernels as its Llama-base ancestor (26.9\% vs. 7.7\%), and on the kernels where it does succeed, it produces designs that are 2.31× faster (geomean) than GPT-5-mini's. When paired with a planner in agentic flow, the \sysname{} codegen outperforms other baselines when optimizing complex code and drives the average normalized improvement across 26 kernels in HLS-Eval suite to \SI{26.4}{\percent}.

\noindent
\textbf{Acknowledgment}: we would like to thank the support from Hewlett Packard Enterprise, NSF (Awards No. 2443992), and AMD.

\bibliographystyle{ACM-Reference-Format}
\bibliography{references}
\appendix

\section{Artifact Appendix}

\subsection{Abstract}

This artifact reproduces the two evaluation workflows reported in the paper: zero-shot HLS pragma generation (Section~4.1) and the iterative agentic flow built on Autocomp (Section~4.2). It provides both training and inference scripts for PRISM .


\subsection{Artifact check-list (meta-information)}

{\small
\begin{itemize}
  \item {\bf Algorithm: } Structure-augmented LLM decoding for HLS pragma
        insertion; AST/CFG/DFG graph encodings prepended as soft-prompt tokens
        to a LoRA-adapted Llama~3.1~8B.
  \item {\bf Program: } Two CLI entry points
        (\texttt{artifact/run\_zero\_shot.py}, \texttt{artifact/run\_agentic.py}),
        a HumanEval benchmark, and a cached-metric figure script.
  \item {\bf Compilation: } AMD/Xilinx Vitis HLS 2023.1 (C~simulation,
        testbench, and synthesis) for the live evaluation paths only.
  \item {\bf Model: } Llama~3.1~8B~Instruct plus a graph-LoRA adapter; OpenAI
        \texttt{gpt-5.4-mini-2026-03-17}; \texttt{gpt\_oss\_20b}.
  \item {\bf Model availability and access: } Base weights are gated under
        Meta's license. The
        adapter is archived at 10.5281/zenodo.21636764. API models are not
        archivable.
  \item {\bf LLM prompts and inference settings: } Prompt construction code and
        per-run prompt dumps are included; inference settings are recorded in
        \texttt{artifact/configs/zero\_shot\_paper.json} and
        \texttt{agentic\_paper.json}. See Section~\ref{sec:ae-models}.
  \item {\bf Data set: } Instruct-Coder(training), HLS-Eval, ForgeBench.
        not redistributed.
  \item {\bf Run-time environment: } Linux, Python~3.11, \texttt{uv} for the
        pinned CUDA environment. Docker image provided for the offline path.
  \item {\bf Container, VM, or locked environment: } \texttt{Dockerfile} based
        on a digest-pinned \texttt{python:3.11-slim}; \texttt{uv.lock} pins the
        full inference environment.
  \item {\bf Hardware: } NVIDIA GPU with $\geq$\,24\,GB VRAM
        ; no GPU required for the offline path.
  \item {\bf Run-time state: } Run directories carry resolved configs, job
        plans, incremental event logs, generated sources, and per-case
        evaluation records. Complete records are reused on restart.
  \item {\bf Execution: } \texttt{docker run} for the offline path;
        \texttt{uv run} with environment variables for model and checkpoint
        paths for live reruns.
  \item {\bf Metrics: } pass@1 over compile, testbench, and synthesis success;
        Vitis latency and LUT/FF/BRAM/DSP/URAM
        utilization.
  \item {\bf Output: } \texttt{results.csv}, \texttt{summary.json},
        \texttt{failures.json}, per-case \texttt{evaluation.json}, generated
        C++, and regenerated figures.
  \item {\bf Expected results and tolerances: } See
        Section~\ref{sec:ae-eval}. Exact LLM text reproduction is not expected
        and is not central to any claim.
  \item {\bf Experiments: } Table~1 (HumanEval), Table~2 (zero-shot), Table~3 (agentic evidence scope), Figure~3
        (metric comparison), Figure~4 (agentic iteration curves).
  \item {\bf Proprietary EDA tools, PDKs, and licensed technology files: }
        AMD/Xilinx Vitis HLS~2023.1. Not included and
        not redistributable. No foundry PDK is required.
  \item {\bf How much disk space required (approximately)?: }
        30\,GB for the offline path, with benchmarks, base
        model, and adapter staged.
  \item {\bf How much time is needed to prepare workflow (approximately)?: }
        10 minutes for the Docker offline path; 1 hour for the
        full environment including model downloads.
  \item {\bf How much time is needed to complete experiments (approximately)?: }
        2-3 days (full reruns).
  \item {\bf API cost or GPU-hours required (if applicable)?: }
        \$20-50 depends on setting.
  \item {\bf Publicly available?: } Yes.
  \item {\bf Code licenses (if publicly available)?: } Apache-2.0.
  \item {\bf Data/model licenses and usage restrictions (if applicable)?: }
        Adapter derives from Llama~3.1 and inherits Meta's license. ForgeBench
        is MIT. The pinned HLS-Eval revision declares no license, so its
        content is fetched rather than redistributed.
  \item {\bf Workflow framework used?: } None; plain Python CLIs with JSON
        manifests.
  \item {\bf Zenodo DOI for archived artifact: } 10.5281/zenodo.21636764
\end{itemize}
}

\subsection{Description}


Source repository: \url{https://github.com/HaochengX/GACO/tree/mlcad-2026-prism-artifact}. Archived record with the adapter bundle:
10.5281/zenodo.21636764. The Zenodo record is the version evaluated.

\subsubsection{Hardware dependencies}

The offline path requires only a Linux host capable of running Docker. Live reruns require a GPU with at least 24\,GB of VRAM.

\subsubsection{Software dependencies}
Check README



\subsubsection{Commercial software and PDK dependencies}

AMD/Xilinx Vitis HLS~2023.1.



\subsubsection{Data sets}

Fetched and checked out at immutable revisions by
\texttt{run\_agentic.py setup}:

{\small
\begin{itemize}
  \item HLS-Eval, \url{https://github.com/sharc-lab/hls-eval}
  \item ForgeBench, \url{https://github.com/hchen799/ForgeBench}
  \item Autocomp, \url{https://github.com/ucb-bar/autocomp}
\end{itemize}
}

\subsubsection{Models}
\label{sec:ae-models}

\textbf{PRISM.} Llama~3.1~8B~Instruct (gated under
Meta's license) with the graph-LoRA adapter \texttt{0910\_layer1}
(633\,MB), archived at 10.5281/zenodo.21636764. 





\subsection{Installation}

{\small
\begin{verbatim}
docker build -t prism-ae .
docker run --rm prism-ae
sudo docker run --rm -it prism-ae bash
python -m pytest tests/artifact -q
bash check_install.sh
\end{verbatim}
}

\noindent The final command runs the offline test suite and is the smoke test.
Expected output: e.g. "N passed, N skipped in Ns".
The repository-hygiene tests skip inside the image because it contains no git
checkout; this is expected. The image requires no network access at run time,
which can be confirmed with \texttt{docker run -{}-rm -{}-network none prism-ae}.



\subsection{Experiment workflow}
\texttt{run\_zero\_shot.py} implements Section~4.1 as independent one-shot
generation: one sample per (model, case), with no planner, search, feedback,
repair, or baseline fallback. The full manifest holds 106 cases; Table~2 uses
the 78-case intersection present in all four executed metrics files.
Generation and evaluation are pipelined --- a case's generated source is
consumed by a Vitis worker as soon as it is written atomically, so generation
does not block on synthesis of the whole suite.

\texttt{run\_agentic.py} implements the iterative Autocomp flow. The
\texttt{figure4} preset covers \texttt{2mm}, \texttt{3mm}, and
\texttt{parallel\_merge\_sort} over ten iterations. The
\texttt{table3-evidence} preset covers the saved evidence scope: five
configurations $\times$ 15 cases $\times$ two repetitions, at three Autocomp
iterations per job, for 150 jobs.

\subsection{Evaluation and expected results}
\label{sec:ae-eval}
From the committed metrics CSVs with no model, API, or synthesis involvement,
\texttt{figure/figure3/process\_results.py} regenerates Figure~3;
\texttt{figure/figure4/plot\_agent.py} regenerates Figure~4.
\texttt{testing/Humaneval\_benchmark.py} reproduce the benchmark results of table 1 for HumanEval.

Table~\ref{tab:ae-claims} maps each paper result to the command that produces
it, the expected output, the similar-result criterion, and the intended
validation path. Paths are: \emph{offline} (no credentials or licensed tools),
\emph{cached} (recomputed from committed records), and \emph{full} (live
models and synthesis).

\begin{table}[t]
  \centering
  \caption{Validation map for each key result.}
  \label{tab:ae-claims}
  \setlength{\tabcolsep}{3pt}
  \small
  \begin{tabular}{llll}
    \toprule
    \textbf{Result} & \textbf{Command} & \textbf{Criterion} & \textbf{Path} \\
    \midrule
    Suite    & \texttt{docker run prism-ae}         & all pass/skip & offline \\
    Fig.~3   & \texttt{process\_results.py}          & plot    & cached  \\
    Fig.~4   & \texttt{run\_agentic.py -{}-preset figure4} & plot & cached \\
    Table~1  & \texttt{Humaneval\_benchmark.py}      & pass rate    & full    \\
    Table~2  & \texttt{run\_zero\_shot.py paper}     & pass rate   & full    \\
    Table~3  & \texttt{-{}-preset table3-evidence}   & pass rate  & full    \\
    \bottomrule
  \end{tabular}
\end{table}




\subsection{Experiment customization}

Individual lanes or cases can be selected by repeating \texttt{-{}-model} or
\texttt{-{}-only}. Summaries can be recomputed from an existing run directory
without re-invoking any model via
\texttt{run\_zero\_shot.py summarize -{}-run-dir <dir>}. Model and adapter
locations are overridden through \texttt{GACO\_MODEL\_PATH} and
\texttt{GACO\_CHECKPOINT\_PATH}. Setting \texttt{PRISM\_AUTOCOMP\_SOURCE} to a
checkout of the pinned Autocomp commit enables offline verification that the
provided patch still applies.

One can also use \texttt{training/train\_fusion\_single.py} to work on a customized adapter. 








\section{Additional Details}
\subsection{Training Details}
\begin{itemize}
\item \textbf{Hardware}: 2$\times$ H100
\item \textbf{Batch Size}: 14
\item \textbf{Gradient accumulation}: 2 (Effective batch size 28)
\item \textbf{Learning rate}: 1 $\times$ $10^{-6}$
\item \textbf{Optimizer}: AdamW, betas=(0.9,0.95), eps=1 $\times$ $10^{-6}$
\item \textbf{Weight Decay}: 0.01
\item \textbf{LR Scheduler}: Cosine Annealing
\item \textbf{Precision}: bfloat16
\item \textbf{LoRA}: rank=12, alpha=16, dropout=0.1
\item \textbf{Target Modules}: q\_proj, k\_proj, v\_proj, o\_proj
\end{itemize}

\subsection{Model Architecture}

\begin{table}[H]
\centering
\begin{tabular}{llll}
\toprule
\textbf{Component} & \textbf{Input Shape} & \textbf{Output Shape} & \textbf{Parameters} \\
\midrule
AST GNN & [x, 128] & [B, 256] & $\sim$200K \\
CFG GNN & [x, 128] & [B, 256] & $\sim$200K \\
DFG GNN & [x, 128] & [B, 256] & $\sim$200K \\
Fusion & [B, 768] & [B, 768] & $\sim$590K \\
TokenGen & [B, 768] & [B, 128, 768] & $\sim$1.2M \\
Graph$\rightarrow$LLaMA & [B, 128, 768] & [B, 128, 4096] & $\sim$3.1M \\
LLaMA+LoRA & [B, L, 4096] & [B, L, vocab] & $\sim$100M (LoRA) \\
\bottomrule
\end{tabular}

\vspace{1em} 
\raggedright
Overall trainable parameters $\sim$106M
\end{table}

\subsection{Limitations} 
First, the graph encoders add per-token inference overhead; the loop-level token efficiency offsets this in the agentic setting, but \sysname{} remains slower per forward pass than the vanilla base model. 
Second, the cross-language transfer from Python-trained graph encoders to C/C++ HLS sources relies on a node-type mapping that is necessarily imperfect (declarations and explicit initializations in C have no Python analogue), and this accounts for a measurable share of the residual failure rate.
Third, \sysname{} requires a complete, parsable kernel: partial code or syntactically broken input cannot be reasoned about structurally, since the graph streams collapse before the gated cross-attention can act.

\subsection{Future directions} 
The most natural extension is to drop the Python detour entirely and build graph encoders over Clang ASTs and LLVM IR for C/C++, thereby exposing pointer aliasing, memory hierarchy, and the explicit type information that the current node-type mapping discards.
The same architecture should transfer to RTL design with Verilog and SystemVerilog by replacing the program graphs with module-hierarchy, signal-dependency, and timing-constraint graphs, and to SystemC by adding process-network and transaction-level representations.
Two questions also remain open: (1) whether the flow we exploit (syntax, control, semantics) is universal across code-transformation tasks or specific to pragma optimization, and (2) whether a GNN surrogate such as HARP can serve as a downstream verifier or reward signal that closes the loop between \sysname{}'s generative front end and the QoR-aware discriminative models the HLS community has spent years building.

\end{document}